\documentclass[prl,twocolumn,groupedaddress]{revtex4}
\usepackage{graphicx}
\usepackage{amsmath}
\usepackage{amssymb}
\usepackage{bm}
\usepackage{color}
\usepackage{hyperref}
\usepackage{tabularx}
\usepackage{multirow}
\usepackage{braket}
\usepackage{epstopdf}

\usepackage{stackengine}
\newcommand\xrowht[2][0]{\addstackgap[.5\dimexpr#2\relax]{\vphantom{#1}}}
\newcolumntype{C}[1]{>{\centering\arraybackslash}m{#1}}

\newcommand{\be}{\begin{equation}}
\newcommand{\ee}{\end{equation}}

\newcommand{\bea}{\begin{eqnarray}}
\newcommand{\eea}{\end{eqnarray}}

\begin{document}
\title{Orbital Fulde-Ferrell State versus Orbital Larkin-Ovchinnikov State} 
	
\author{Noah F. Q. Yuan}
\email{fyuanaa@connect.ust.hk}
\affiliation{Tsung-Dao Lee Institute, Shanghai Jiao Tong University, Shanghai 201210, China}
\affiliation{School of Physics and Astronomy, Shanghai Jiao Tong University, Shanghai 200240, China}

\begin{abstract}
Orbital Fulde-Ferrell-Larkin-Ovchinnikov (FFLO) states are analyzed within mean-field theory, where orbital FF state is layer-polarized with inversion symmetry broken, while orbital LO state is Josephson vortex array with translation symmetry reduced. 
Phase diagrams of orbital FFLO states are obtained, and properties such as induced orders, superconducting diode effects, Fraunhofer pattern and topological defects are studied for the probe of FF versus LO states.
\end{abstract}
                          	
\maketitle
\textit{\textcolor{blue}{Introduction}}---
In conventional superconductors, Cooper pairs are formed with zero momentum at zero field. 
At high magnetic fields and low temperatures, Cooper pairs may carry finite momentum in the Fulde-Ferrell-Larkin-Ovchinnikov (FFLO) states \cite{FF,LO}, which can be induced by the Zeeman effect of magnetic fields. 

In a two-dimensional (2D) Ising superconductor \cite{JMLu,XXi,YuS}, Cooper pairs could survive under high in-plane magnetic fields, as the Ising spin-orbit coupling (SOC) suppresses the Zeeman effect of in-plane fields. 
In this case one may consider the role of orbital effect.
It has been proposed that the orbital effect of in-plane fields can induce finite-momentum Cooper pairs in bilayer Ising superconductors \cite{CXL}, whose evidence was later reported in Ising superconductor few layers \cite{OFFLO1a}, thin flakes \cite{OFFLO1,OFFLO1b,OFFLO1e} and even the bulk \cite{OFFLO1c,OFFLO1d}. 
Similar to the FFLO states but induced by the orbital effect, such finite-momentum pairing state is called the orbital FFLO states \cite{OFFLO2,OFFLO3}.


However, the spatial configurations of the order parameters in orbital FFLO states remain under debates. 
In the case of a bilayer Ising superconductor under an in-plane field, 
in Refs. \cite{OFFLO2,OFFLO3,CXL} order parameters in two layers are both uniform plane waves with the same \cite{OFFLO2,OFFLO3} or opposite \cite{CXL} Cooper pair momenta,
while in Refs. \cite{GWQ,UN,HY}, order parameters are spatially modulated.

In this work, we specifically study a model of bilayer Ising superconductors, and carry out controllable mean-field calculations to determine the spatial configurations of the orbital FFLO states. We first work out possible phases with the help of symmetry analysis, then obtain the temperature-field phase diagram. Finally we propose experimental probes such as superconducting diode effects to detect orbital FFLO states. 
We also envision the orbital FFLO states in multilayer and bulk cases.


\begin{figure}
\includegraphics[width=\columnwidth]{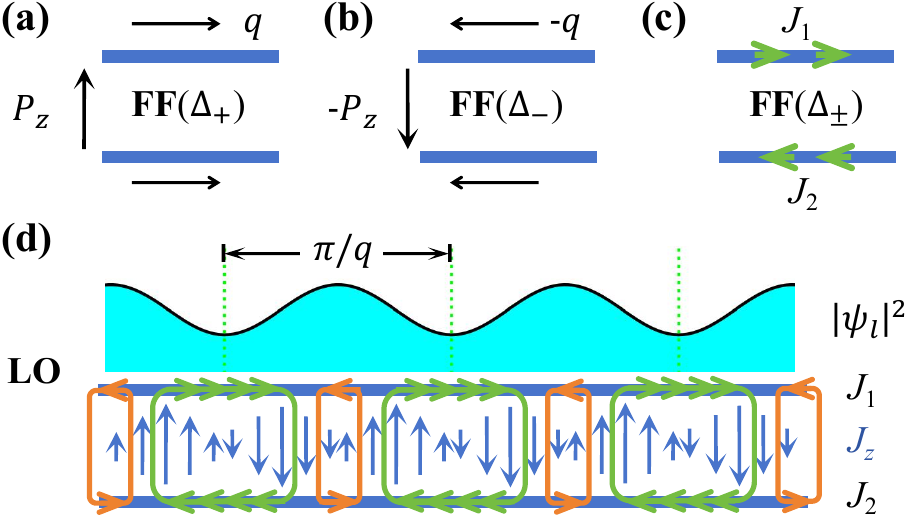}
\caption{Schematics of the orbital FF (a-c) and LO (d) states. Two types of orbital FF states have opposite Cooper pair momenta $\pm q$ and electric polarizations $\pm P_z$ (a,b), but the same supercurrent distribution $\bm J_{1,2}$ (c). (d) The orbital LO state is an array of Josephson vortices (green) and antivortices (orange), hence the observables $|\psi_l|^2,\bm J_{1,2},J_z$ are modulated with periodicity $\pi/q$. Details can be found in the maintext.}\label{fig1}
\end{figure}


\textit{\textcolor{blue}{Model}}---
We consider a bilayer Ising superconductor whose Ising SOC supresses the Zeeman effect of an in-plane magnetic field $\bm B$. 
In our model, we ignore the Zeeman effect and the free energy density is
\bea\label{eq_f1}
f&=&f_{\bm Q}[\psi_1]+f_{-\bm Q}[\psi_2]+f_c[\psi_1,\psi_2],\\
f_{\bm Q}[\psi]&=&\frac{1}{2m}\left|\left(\nabla-i\bm Q\right)\psi\right|^2+a|\psi|^2+b |\psi|^4,\\
f_c&=&\mathcal{J}(\psi_1^*\psi_2+c.c.)+{\mathcal{K}}[(\psi_1^*\psi_2)^2+c.c.].
\eea
Here $\psi_{l}=\psi_{l}(\bm r)$ is the local order parameter on layer $l=1,2$ at in-plane position $\bm r$, 
and the orbital effect is described by the orbital field in a Landau gauge,
\be\label{eq_A}
\bm Q=\frac{\pi d}{\Phi_0}\bm B\times\hat{\bm z}
\ee
with flux quantum $\Phi_0$, interlayer distance $d$,
conventional Ginzburg-Landau parameters $m,a,b$, 
and Josephson coupling parameters $\mathcal{J},\mathcal{K}$.
Without loss of generality, we assume $\mathcal{J}<0$ in this manuscript, and the spatially uniform phase $\psi_1=\psi_2$ is favored at zero field, which is zero-momentum pairing.
The expansion of free energy density is up to the \textit{fourth} order of order parameters.



The symmetry group of free energy density Eq. (\ref{eq_f1}) is 
$G={\rm U}(1)\times\mathbb{R}^2\rtimes D_{2h}$.
The continuous part ${\rm U}(1)\times\mathbb{R}^2$ includes U(1) gauge transform $U_{\omega}$ and in-plane translation $T_{\bm a}$ with $\omega\in[0,2\pi],\bm a\in\mathbb{R}^2$.
The discrete part $D_{2h}$ is generated by spatial inversion $I$, vertical mirror 
$M:\bm v\to\bm v-2\hat{\bm B}(\bm v\cdot\hat{\bm B})$, and the antiunitary twofold in-plane rotation $C_2\mathcal{T}$,
although both twofold rotation $C_2$ and time-reversal $\mathcal{T}$ are broken by the magnetic field. 
Representation of $G$ on $\psi_l(\bm r)$ is listed in Tab. \ref{tab_1}.


In our model, we neglect Zeeman effect and crystal anisotropy. 
To include weak Zeeman effect, we allow $a$ to have field dependence $a=a_0+a_1B^2$ with Zeeman effect coefficient $a_1>0$.
To describe the crystal anisotropy, we need sixth order terms in free energy \cite{OFFLO2}, and the symmetry group reduces to $G'={\rm U}(1)\times\mathbb{R}^2\rtimes \mathbb{Z}_2$, where $C_2\mathcal{T},M$ are broken and $\mathbb{Z}_2$ is generated by inversion $I$.

Phases in our model should be found by minimizing the free energy, usually by numerical methods \cite{OFFLO3,GWQ,UN,HY}. 
In Ref. \cite{OFFLO2}, we consider the vicinity of phase transitions and find out possible phases analytically by neglecting the fourth order terms in free energy.
We find as the field increases, the Cooper pair momentum remains zero until the field is stronger than a critical value, 
where superconducting states become the so-called orbital FFLO states.
In this work, we include the fourth order terms and compute the full free energy of the orbital FFLO states to figure out the stable phase. 

\textit{\textcolor{blue}{Phases}}---
We first neglect the fourth order terms, then
we obtain the following orbital FFLO ansatz 
\be\label{eq_fflo}
\begin{pmatrix}
    \psi_1\\
    \psi_2
\end{pmatrix}
=\Delta_{+}e^{i\bm q\cdot\bm r}
\begin{pmatrix}
    \cos\vartheta\\
    \sin\vartheta
\end{pmatrix}
+\Delta_{-}e^{-i\bm q\cdot\bm r}
\begin{pmatrix}
    \sin\vartheta\\
    \cos\vartheta
\end{pmatrix},
\ee
with parameters $\Delta_{\pm},\bm q$ and $\vartheta$ to be determined.
The complex order parameters $\Delta_{\pm}$ turn out to furnish a two-dimensional (2D) irreducible representation of $G$, labeled by Cooper pair momentum $\bm q\parallel\bm Q$, as listed in Tab. \ref{tab_1}.
Up to the second order terms, the free energy of the orbital FFLO ansatz Eq. (\ref{eq_fflo}) has emergent symmetry group SU(2) acting on $(\Delta_{+},\Delta_{-})$, which is larger than $G$.

Next we consider the full free energy density ${f}$ to determine $\Delta_{\pm}$ configuration. 
Up to the fourth order of $\Delta_{\pm}$, three linearly independent invariants are found under $G$ according to Tab. \ref{tab_1}. As a result the free energy $F\equiv\int fd^2\bm r$ as a quartic function of $\Delta_{\pm}$ can be obtained by a linear combination of these three invariants 
\cite{OFFLO2}
\bea\label{eq_o2}
F=\alpha(|\Delta_{+}|^2+|\Delta_{-}|^2)+{\beta}_{+}(|\Delta_{+}|^2+|\Delta_{-}|^2)^2\\\nonumber
+ {\beta}_{-}(|\Delta_{+}|^2-|\Delta_{-}|^2)^2
\eea
with coefficients $\alpha$ and ${\beta}_{\pm}$ calculated in the Appendix.



When ${\beta}_{-}<0$, free energy Eq. (\ref{eq_o2}) is minimized by the orbital FF phase $\Delta_{+}\Delta_{-}=0$ and the 
inversion symmetry is spontaneously broken, as depicted in Fig. \ref{fig1}(a-c). 

When ${\beta}_{-}>0$, free energy Eq. (\ref{eq_o2}) is minimized by the orbital LO phase $|\Delta_{+}|=|\Delta_{-}|$ and 
translation symmetry is spontaneously reduced, as depicted in Fig. \ref{fig1}(d).

\begin{table}[t]
\centering
\begin{center}  
\begin{tabular}{C{0.8cm}|C{1.5cm}|C{1.5cm}|C{0.8cm}|C{0.8cm}||c|c|c|c}  
\hline
 & $\psi_l(\bm r)$ & $\Delta_{\pm}$ & $\bm P_{\parallel}$ & $P_z$ & N & U & FF & LO \\
\hline\xrowht{10pt}
 {$U_{\omega}$} & 
$\psi_{l}(\bm r)e^{i\omega}$ & $\Delta_{\pm}e^{i\omega}$ & $\bm P_{\parallel}$ & $P_z$ & $\checkmark$ & $\times$ & $\times$ & $\times$\\
\hline\xrowht{10pt}
 {$T_{\bm a}$} & 
$\psi_{l}(\bm r+\bm a)$ & $\Delta_{\pm}e^{\pm i\bm q\cdot\bm a}$ & $\bm P_{\parallel}$ & $P_z$ & $\checkmark$ & $\checkmark$ & $\checkmark$ & $\times$\\
\hline\xrowht{10pt}
 {$I$} & 
$\psi_{\overline{l}}(-\bm r)$ & $\Delta_{\mp}$ & $-\bm P_{\parallel}$ & $-P_z$ & $\checkmark$ & $\checkmark$ & $\times$ & $\checkmark$\\
\hline\xrowht{10pt}
{$C_2\mathcal{T}$} & $\psi_{l}^*(-\bm r)$ & $\Delta_{\pm}^*$ & $-\bm P_{\parallel}$ & $P_z$ & $\checkmark$ & $\checkmark$ & $\checkmark$ & $\checkmark$\\
\hline
\end{tabular}  
\end{center} 
\caption{\textbf{Representations of symmetry group.} Here $l=1,2$ and $\overline{l}=3-l$ is opposite to $l$, and $\bm P_{\parallel}=(P_x,P_y)$. 
$\checkmark$ or $\times$ means the phase is invariant or not under given symmetry.
}
\label{tab_1}
\end{table}




Up to now we choose a specific Landau gauge for the orbital effect,
which induces nonzero Cooper pair momentum $\bm q$.
However, $\bm q$ is gauge-dependent and not an observable after all. 
We thus work out the thermodynamic conjugate of $\bm q$, the supercurrent density, as the observable to characterize orbital FFLO states.
The in-plane supercurrent density of layer $l$ is
\be
\bm J_l=\frac{2e}{m}{\rm Re}\left\{\psi_l^*[-i\nabla+(-1)^l\bm Q]\psi_l\right\},
\ee
and the out-of-plane supercurrent density $J_z$ reads
\be\label{eq_Jz}
J_z=
ed\mathcal{J}{\rm Im}(\psi_1^*\psi_2),
\ee
which is nothing but the Josephson current density.



In orbital FF state, $\bm J_{1,2}$ are spatially uniform and $J_z\equiv 0$,
as depicted in Fig. \ref{fig1}(c). 
In orbital LO state, 
as shown in Fig. \ref{fig1}(d), the order parameter amplitude squared $|\psi_l|^2$, in-plane supercurrent 
$\bm J_{l}$
and Josephson current $J_z$
are all spatially modulated with wavevector $2\bm q$. 
Due to the preserved inversion symmetry, $|\psi_1|^2=|\psi_2|^2$ and $\bm J_1=-\bm J_2$. The extrema points of $|\psi_l|^2$ and $\bm J_{l}$ are the same, which coincide with the nodal points of $J_z$. 


We have distinguished orbital FF and LO states by their supercurrent distributions. However, two types of orbital FF states ($\Delta_{+}$ versus $\Delta_{-}$) share the same supercurrent distribution, and we need a new observable, the electric polarization $\bm P$. 
As shown in Tab. \ref{tab_1}, the coupling between $\Delta_{\pm}$ and $\bm P$ is described by the free energy
$F_{\rm c}\propto P_z(|\Delta_{+}|^2-|\Delta_{-}|^2)$.
By minimizing the total free energy we find an out-of-plane electric polarization 
$\bm P\propto(|\Delta_{+}|^2-|\Delta_{-}|^2)\hat{\bm z}$,
which is opposite for two types of orbital FF states $\Delta_{\pm}$ as shown in Fig. \ref{fig1}(a,b). 
Reversely, applying a vertical gate voltage could stabilize FF state over LO state \cite{OFFLO3}.
Notice that the emergent symmetry $C_2\mathcal{T}$ pins $\bm P$ to $z$-direction, and by taking into account crystal anisotropy, the induced electric polarization in FF phase can also have in-plane components. 
Similarly, charge (or spin) density wave will be induced with wave vector $2\bm q$ in the orbital LO state.

To summarize, we find that orbital FF state is in-plane uniform and layer-polarized [Fig. \ref{fig1}(a,b)], while orbital LO state is an array of Josephson vortices (green) and antivortices (orange) [Fig. \ref{fig1}(c)]. Next we will figure out the conditions for orbital FF/LO states.

\textit{\textcolor{blue}{Phase diagram}}---
In our model we can find four phases by symmetry analysis, namely, the orbital FF phase, the orbital LO phase, the spatially uniform phase (U) and the normal phase (N), whose symmetries are summarized in Tab. \ref{tab_1}.
Among them, N furnishes the trivial representation, U furnishes a 1D irreducible representation, and FFLO states furnish a 2D irreducible representation.
The irreducible representations of $G$ can also be 4D, whose free energy Eq. (\ref{eq_f1}) are found higher than that of any 1D or 2D representation, and hence are not considered here. 

\begin{figure}
\includegraphics[width=\columnwidth]{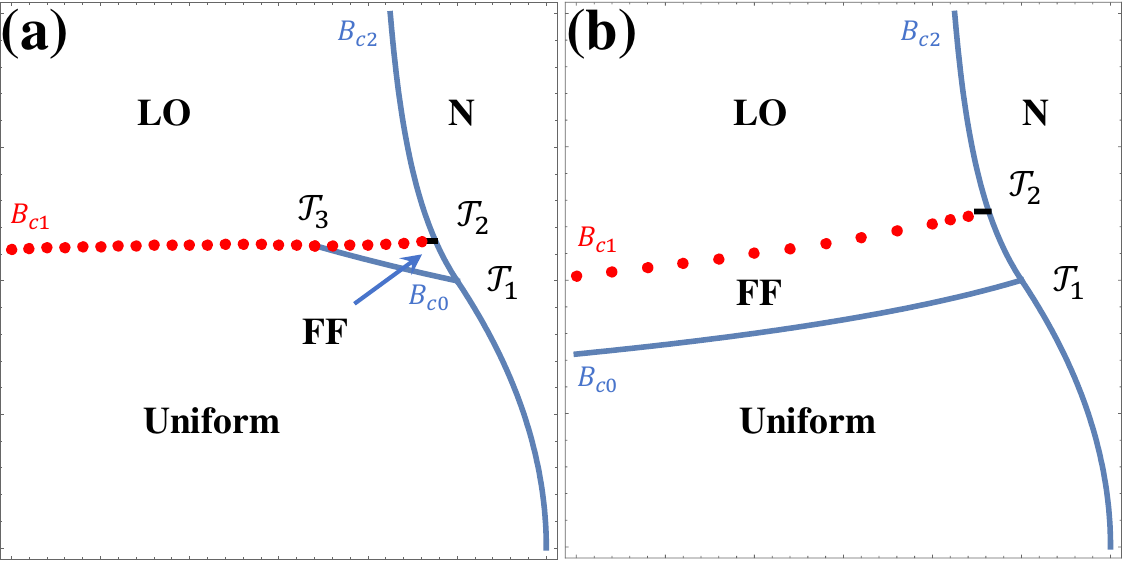}
\caption{Mean-field phase diagrams of the bilayer model
Eq. (\ref{eq_f1}), with four phases N, U, FF, LO, and tricritical points $\mathcal{T}_{1,2,3}$, as defined in the maintext. 
We set $\kappa=1/2,2$ for (a,b) respectively.
The horizontal axis is reduced temperature $t$, the vertical axis is reduced field $h$, as defined in Eq. (\ref{eq_p}). Solid lines denote the 2nd-order phase transitions while red dotted lines denote the 1st-order phase transitions.}\label{fig2}
\end{figure}

Our model Eq. (\ref{eq_f1}) has three dimensionless parameters 
\be\label{eq_p}
t=a/|\mathcal{J}|,\quad h=Q/\sqrt{m|\mathcal{J}|},\quad
\kappa=\mathcal{K}/b. 
\ee
At given reduced temperature $(t)$ and reduced field $(h)$ one can derive the stable phase by (numerically) minimizing the free energy in Eq. (\ref{eq_o2}). 
Then the $t$-$h$ phase diagram is obtained at given $\kappa$, as shown in Fig. \ref{fig2}.

Tricritical points can be found in the phase diagram.
At the \textit{first} tricritical point $\mathcal{T}_1$, U, FF and N coexist and 
at the \textit{second} tricritical point $\mathcal{T}_{2}$, FF, LO and N coexist.

When $-1<\kappa<\kappa_{c}\approx 1$, besides tricritical points $\mathcal{T}_{1,2}$ one finds the \textit{third} tricritical point $\mathcal{T}_{3}$, 
where U, FF and LO coexist, as shown in Fig. \ref{fig2}(a) with $\kappa=1/2$.
The orbital FF phase is confined in the delta region $\mathcal{T}_1\mathcal{T}_2\mathcal{T}_3$, which expands with increasing $\kappa$.
Ising superconductor thin films in Ref. \cite{OFFLO1b} may belong to this case.

When $\kappa\geq\kappa_c$, there are two tricritical points $\mathcal{T}_{1,2}$, as in Fig. \ref{fig2}(b) with $\kappa=2$.
The uniform phase and orbital LO phase are separated by orbital FF phase in between. 
As $\kappa$ increases, the belt region of orbital FF state in the phase diagram expands.
Ising superconductor thin films in Refs. \cite{OFFLO1,OFFLO1q} may belong to this case.




The phase transitions between superconducting phases (U, FF and LO) and the normal phase (N) are 2nd-order, 
leading to the in-plane \textit{upper} critical field $B_{c2}$ \cite{OFFLO2}, as shown in Fig. \ref{fig2}. 
When $B=B_{c2}$ we have $\cot 2\vartheta=\lambda^2\bm q\cdot\bm Q$, 
\be\label{eq_q0b}
\bm q=\bm Q{\rm Re}\sqrt{1-\left(\frac{B_*}{B}\right)^4},\
{\beta}_{-}=\frac{b}{2}\left\{\left(\frac{B_{**}}{B}\right)^4-1\right\},
\ee
where $B_*$ and $B_{**}$ are corresponding fields of $\mathcal{T}_1$ and $\mathcal{T}_2$ 
\be\label{eq_B*}
B_*=\frac{\Phi_0}{\pi\lambda d},\quad
B_{**}=B_{*}\left(\frac{3+\kappa}{2}\right)^{1/4},
\ee
with $\lambda=1/\sqrt{m|\mathcal{J}|}$ and $\kappa=\mathcal{K}/b$. 
To ensure the stability of the free energy density in Eq. (\ref{eq_f1}), we require $m,b>0$ and $\mathcal{K}+b>0$.
Hence $\kappa>-1$, and $B_{**}>B_{*}$. In other words, the orbital FF phase could exist in the field range $B_*<B<B_{**}$ as long as the system is stable.

The phase transition from LO to FF (or U) is the Josephson vortex array melting, 
which is 1st-order (red dots in Fig. \ref{fig2}) within mean-field theory \cite{GWQ,HY}. We may denote the transition field in this 1st-order phase transition as the in-plane \textit{lower} critical field $B_{c1}$.
In-plane fields are screened when $B<B_{c1}$, and penetrate to create Josephson vortex array when $B>B_{c1}$.

The phase transition between U and FF is 2nd-order, which may introduce the in-plane \textit{lowest} critical field $B_{c0}$. We find $B_{c0}\leq B_{c1}\leq B_{c2}$ at the same temperature, and $B_{c0}=B_{c2}=B_*$ at $\mathcal{T}_1$, $B_{c1}=B_{c2}=B_{**}$ at $\mathcal{T}_2$, and $B_{c0}=B_{c1}$ at $\mathcal{T}_3$, as shown in Fig. \ref{fig2}.

As the field increases, the translation symmetry will be spontaneously broken, while the inversion symmetry can change from preserved (U) to broken (FF), and then back to restored (LO). 
To probe orbital FF state, we keep track of the field evolution of inversion symmetry by analyzing diode effects \cite{Ando,Lorenz,Ilic,Yuan,Chris,Diez,LinJ,Harley,Akito,James,Zhai,Marg,Banabir,JXHu,Samokhin,James1,OFFLO3,OFFLO4} of in-plane current. 

\begin{figure}
\includegraphics[width=\columnwidth]{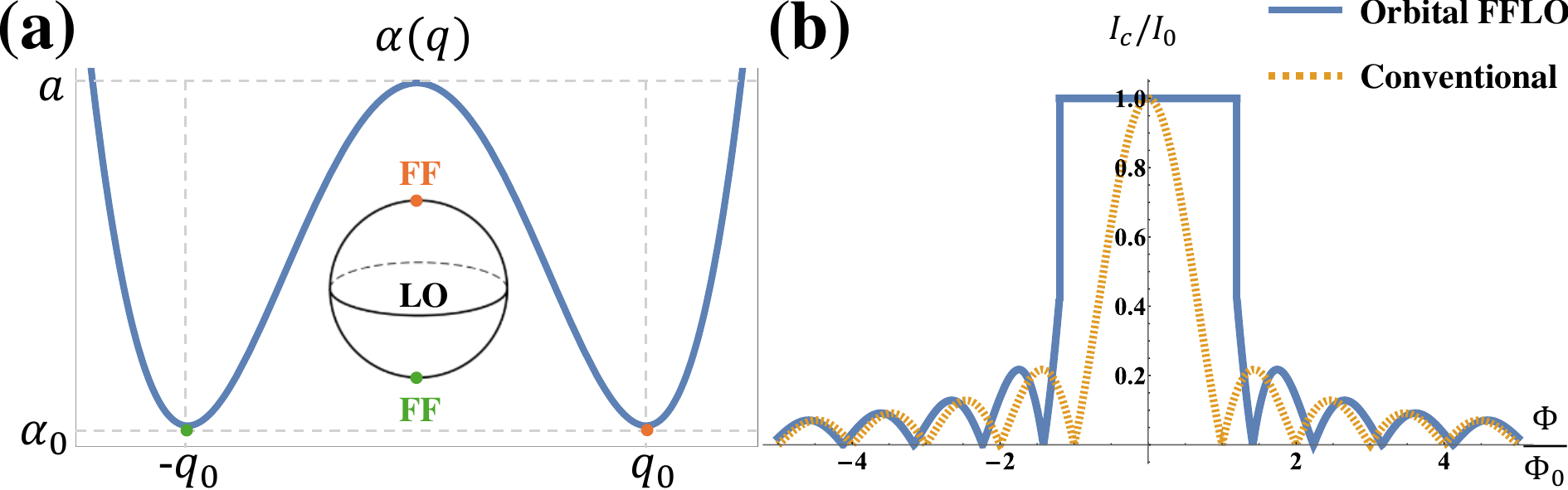}
\caption{(a) Plot of $\alpha(\bm q)$ in Eq. (\ref{eq_o2}) when $\bm q\parallel\bm Q$, with $a$ defined in Eq. (\ref{eq_f1}) and $\alpha_0$ defined in Eq. (\ref{eq_f2}). Inset: Bloch sphere of orbital FFLO states. (b) Fraunhofer patterns in orbital FFLO case with $\Phi_{c0}=\Phi_0$, $B_{c1}=1.2B_{c0}$ (blue) and in conventional case (orange), where $\Phi_{c0}$ is the flux of $B_{c0}$.}\label{fig3}
\end{figure}

\textit{\textcolor{blue}{Superconducting diode effects}}---
In the orbital FF state $B_{c0}<B<B_{c1}$, 
the Josephson current is zero $J_z=0$, and the total in-plane supercurrent $\bm J$ is spatially uniform
\be\label{eq_Jt}
\bm J=\bm J_1+\bm J_2=2e|\Delta|^2\partial_{\bm q}\alpha,
\ee
where $\alpha$ is defined in Eq. (\ref{eq_o2}) and plotted in Fig. \ref{fig3}(a),
and $\Delta$ is the order parameter of orbital FF phase (namely $\Delta=\Delta_{+}$ or $\Delta_{-}$).
Without external current source $\bm J=\bm 0$, the equilibrium Cooper pair momentum can be $\pm\bm q_0$, corresponding to FF phase $\Delta_{\pm}$ respectively. 
The expression of $\bm q_0$ is given in Eq. (\ref{eq_q0b}) when $B=B_{c2}$.
Under nonzero supercurrent, $\bm q$ deviates from $\pm\bm q_0$, we can expand 
\bea\label{eq_f2}
\alpha(\bm p\pm\bm q_{0})=\alpha_{0}+\alpha_{2\parallel} p_{\parallel}^2+\alpha_{2\perp} p_{\perp}^2
\pm\alpha_{3} p_{\parallel}^3
\eea
up to the third order
with $\alpha_0=\alpha(\bm q_0)$, $p_{\parallel}=\bm p\cdot\hat{\bm q}_0$, and $p_{\perp}=\bm p\cdot(\hat{\bm z}\times\hat{\bm q}_0)$. 
The condition of $\mathcal{T}_1$ is $\alpha_{2\parallel}=\alpha_0=0$.
The third order coefficient $\alpha_3$ describes the asymmetry of $\alpha(\bm q)$ with respect to $\bm q=\bm q_0$ as shown in Fig. \ref{fig3}(a). This asymmetry indicates the nonreciprocal supercurrent transport related to all three types of superconducting diode effects discussed in the following. 

Beneath the mean-field phase transition ($\alpha_0\leq 0$), the supercurrent diode effect (SDE) is found \cite{Yuan,Akito,James,Harley,Zhai,Ilic}, and the critical current is nonreciprocal under field
\be\label{eq_FF}
\eta\equiv\frac{J_{c}^+-J_{c}^-}{J_{c}^++J_{c}^-} =\sqrt{\frac{|\alpha_0|}{3\alpha_{2\varphi}^3}}\alpha_{3}\cos^3\varphi,
\ee
where $\varphi$ is the angle between supercurrent $\bm J$ and $\bm q_0$, $J_{c}^{\pm}$ are critical currents along $\pm\bm J$ directions respectively, and $\alpha_{2\varphi}=\alpha_{2\parallel}\cos^2\varphi+\alpha_{2\perp}\sin^2\varphi$. 
Below the mean-field phase transition ($\alpha_0<0$), the resistance can still be nonzero due to the proliferation of vortex-antivortex pairs in 2D. One can define the Berezinskii–Kosterlitz–Thouless (BKT) transition temperature $T_{\rm BKT}=\frac{\pi}{2}\sqrt{\det\rho}$ in terms of the superfluid stiffness tensor $\rho_{ij}=\partial^2 F/\partial q_i\partial q_j$. 
Similarly, we can introduce the BKT diode effect with nonreiprocal transition temperature under current, whose coefficient is 
\be\label{eq_BKT}
\zeta\equiv\frac{T_{\rm BKT}^{+}-T_{\rm BKT}^{-}}{T_{\rm BKT}^{+}+T_{\rm BKT}^{-}} =
\frac{3\beta}{2e}
\frac{\alpha_{3}}{|\alpha_0|\alpha_{2\parallel}^2}J\cos\varphi,
\ee
where $T_{\rm BKT}^{\pm}$ are BKT temperatures under supercurrent $\pm\bm J$ respectively, $\beta=\beta_{+}+\beta_{-}$ and $J=|\bm J|$.

Above the mean-field phase transition ($\alpha_0\geq 0$) is the fluctuating regime of superconductivity, where the current $\bm J$ is dissipative and depends on the applied electric field $\bm E$. 
From the expansion Eq. (\ref{eq_f2}), we compute $\bm J\propto\partial_{\bm q}\alpha$ and replace $\bm q$ by $\bm E$, then obtain the current-field relation up to the second order of electric field 
\be\label{eq_je}
J_{\parallel}=\sigma_{\parallel} E_{\parallel}+\chi E_{\parallel}^2,\quad
J_{\perp}=\sigma_{\perp} E_{\perp},
\ee
where $\sigma_{\parallel,\perp}$ and $\chi$ are paraconductivity parameters. 
Notice that in $J_{\parallel}$, besides the conventional linear term $\sigma_{\parallel}$, there is also a nonlinear term $\chi$ known as the nonlinear paraconductivity (NLP), which is absent in $J_{\perp}$.
Within the Langevin theory of fluctuations, we can work out 
\be\label{eq_R2}
\chi=
\frac{e^3\gamma^2 T}{2\pi}\frac{\alpha_{3}}{|\alpha_0|^2\sqrt{\alpha_{2\parallel}\alpha_{2\perp}}},
\ee
where $\gamma>0$ is the inverse of damping constant \cite{James,Sch} and $T$ is the temperature.
From Eq. (\ref{eq_je}), the field-current relations up to the second order of current can be worked out
$E_{\parallel}={J_{\parallel}}/{\sigma_{\parallel}}-{\chi J_{\parallel}^2}/{\sigma_{\parallel}^3},\quad
E_{\perp}={J_{\perp}}/{\sigma_{\perp}}$.

To summarize, three types of superconducting diode effects are found in orbital FF state, namely SDE in Eq. (\ref{eq_FF}), BKT diode effect in Eq. (\ref{eq_BKT}) and NLP in in Eq. (\ref{eq_R2}).
These diode effects are optimal (vanishing) when the supercurrent is parallel (perpendicular) to the zero-current Cooper pair momentum, 
enhanced by superfluid skewness ($\alpha_3$) and prevented by superfluid stiffness ($\alpha_{2\parallel}$).
We can use $\alpha_0=0$ to define the mean-field critical temperature $T_c$ and $\alpha_0\propto T-T_c$. 
Then, SDE is due to mean-field superconductivity and hence has the typical square-root temperature dependence $\eta\propto\sqrt{T_c-T}$, 
while BKT diode effect and NLP are both due to fluctuating superconductivity and hence are negatively correlated to $T_c-T$, namely $\zeta\propto(T_c-T)^{1/2}$ and $\chi\propto(T_c-T)^{2}$.

Such diode effect signals should be found in the orbital FF phase but not the other three phases (N, U and LO).
Calculation details can be found in the Appendix.

We have considered the field evolution of in-plane current in the bilayer superconductor.
In the following, we study the field evolution of out-of-plane current.

\textcolor{blue}{\textit{Fraunhofer pattern}}---
With out-of-plane current density $J_z$ in Eq. (\ref{eq_Jz}), we can calculate the Josephson current $I_z=\int d^2\bm r J_z$, leading to the 
critical Josephson current $I_c=\max |I_z|$ of the bilayer superconductor
\be\label{eq_JJ}
I_c=I_0\left|{\rm sinc}\left(\nu\pi\frac{\Phi}{\Phi_0}\right)\right|,\quad\nu=\frac{q}{Q}\theta(B-B_{c1})
\ee
with 
${\rm sinc}x=\sin x/x$, 
magnetic flux $\Phi$, and Cooper pair momentum $q$.
In both FF and U phases, $\nu=0$, and in LO phase $\nu=q/Q$.
An example of 
the modified Fraunhofer pattern is shown in Fig. \ref{fig3}(b),
whose nodes at weak fields can be significantly shifted from $n\Phi_0$ with integer $n$. 
At high fields, the modified Fraunhofer pattern tends to become the conventional one where the in-plane field does not drive phase transitions in superconductors.

Up to now we consider clean samples without defects. In the following we will briefly discuss topological defects in orbital FFLO states, with details left in the Appendix.

\textit{\textcolor{blue}{Topological defects}}---
The orbital FFLO states 
can be described by the Bloch sphere as shown in Fig. \ref{fig3}(a) inset,
$(\Delta_{+},\Delta_{-})=\Delta e^{i\chi}\left(e^{-i\phi/2}\cos\frac{\theta}{2},e^{i\phi/2}\sin\frac{\theta}{2}\right)$ with $\Delta>0$.
Then FF phase is one of the poles ($\theta=0$ or $\pi$) and LO phase is one point at the equator ($\theta=\pi/2$), with induced electric polarization $\bm P\propto \Delta^2\cos\theta\hat{\bm z}$ and induced density wave Fourier component $\rho_{2\bm q}\propto \Delta^2 e^{i\phi}$ respectively.

Without defects, $\chi,\theta,\phi$ are spatially constant.
The Abrikosov vortices can be described by spatially dependent $\chi(\bm r)$ with integer winding number as the topological invariant.
Besides, in FF phase there can be two types of domains and domain walls, while in LO phase there can be dislocations of density waves, which can be described by $\theta(\bm r)$ and $\phi(\bm r)$ respectively. 
When $\theta(\bm r)$ and $\phi(\bm r)$ are both spatially varying, superconducting skyrmions can be formed.
The topological invariants of vortices, dislocations and skyrmions can be found in the Appendix. 

\textit{\textcolor{blue}{Conclusion}}---
In this work, we study the bilayer Ising superconductor under an in-plane magnetic field whose Zeeman effect is neglected. 
We find four possible phases (FF, LO, U, and N) and at most three tricritical points.
Near the first tricritical point $\mathcal{T}_{1}$ (U-FF-N), the superconducting diode effects are expected to be optimal.
Near the second tricritical point $\mathcal{T}_{2}$ (FF-LO-N), the symmetry group is SU(2) and skyrmions are expected \cite{Dimi,book}.

In the bulk limit, the Josephson vortex array of the orbital LO state would evolve into a Josephson vortex lattice \cite{HY,OFFLO1c,OFFLO1d,JWang}, and the vortex lattice melting process may be realized as a BKT-like transition.

Our theory of orbital FFLO states may apply to Ising superconductors \cite{OFFLO1,OFFLO1a,OFFLO1b,OFFLO1c,OFFLO1d,OFFLO1e}, tilted Ising superconductors \cite{DR,DR1,DR2} and moir\'e Ising superconductors \cite{OFFLO3,KFMak,CoryDean}. 
In fact, given the in-plane paramagnetic limiting field $B_p$, when
\be\label{eq_I}
B_{*},B_{**}\ll B_p,
\ee
our theory could apply in the clean limit. In order to be measurable, $B_{*},B_{**}$ should also be large enough compared with experimental precision. As a result, we would expect the material candidates to have high $B_p$ in the clean limit for the detection of orbital FFLO states.



\textcolor{blue}{\textit{Acknowledgments}}---
The author thanks Puhua Wan and Jianting Ye for inspiring discussions.
The author thanks James Jun He for instructions on calculations of paraconductivity and Xin Liu for discussions on modified Fraunhofer pattern.
This work is supported by the National Natural Science Foundation of China (Grant No. 12174021).

\appendix
\section{Appendix A: Coefficients $\alpha$ and $\beta_{\pm}$}
By straight forward calculations we have
\bea\label{eq_alpha}
\alpha=a+\frac{q^2+Q^2}{2m}-\frac{\bm q\cdot{\bm Q}}{m}\cos 2\vartheta+\mathcal{J}\sin 2\vartheta,\\\label{eq_beta}
{\beta}_{\pm}=\frac{1}{2}\left\{b+\left[\frac{1}{2}(\mathcal{K}-b)\pm (\mathcal{K}+b)\right]\sin^2 2\vartheta\right\}.
\eea
By minimizing $\alpha$ with respect to $\vartheta$ we have the optimal condition $\cot 2\vartheta=\lambda^2\bm q\cdot\bm Q$ with $\lambda=1/\sqrt{m|\mathcal{J}|}$, and
\bea
\alpha(\bm q)=a+\frac{q^2+Q^2}{2m}-|\mathcal{J}|\sqrt{1+\lambda^4(\bm q\cdot{\bm Q})^2},\\
{\beta}_{\pm}(\bm q)=\frac{1}{2}\left\{b+\frac{\frac{1}{2}(\mathcal{K}-b)\pm (\mathcal{K}+b)}{1+\lambda^4(\bm q\cdot{\bm Q})^2}\right\}.
\eea

In the orbital FF (LO) phase, the free energy is
\be
\Omega_{\rm FF}=\alpha\Delta^2+({\beta}_{+}+{\beta}_{-})\Delta^4,\quad
\Omega_{\rm LO}=\alpha\Delta^2+{\beta}_{+}\Delta^4,
\ee
where $\Delta=\sqrt{|\Delta_{+}|^2+|\Delta_{-}|^2}\geq 0$.
The optimal free energy of the superconducting phase can be written as
\be
\Omega_s=-\frac{1}{4}\Gamma^2,\quad\Gamma=\frac{\alpha}{\sqrt{{\beta}_{+}+{\beta}_{-}\theta(-{\beta}_{-})}},
\ee
where the step function $\theta(x)=1$ when $x>0$ and $\theta(x)=0$ when $x<0$.
We then minimize $\Gamma$ to obtain the optimal superconducting phase and corresponding Cooper pair momentum $\bm q_0$. Notice that $\pm\bm q_0$ are both optimal momenta corresponding to the same free energy since $\alpha,{\beta}_{+}$ and ${\beta}_{-}$ are all even in $\bm q$.


Next we consider the following expression
\bea
\alpha(\bm p+\bm q_0)=a+\frac{(p_{\parallel}+q_0)^2+p_{\perp}^2+Q^2}{2m}\\
-|\mathcal{J}|\sqrt{1+\lambda^4Q^2(p_{\parallel}+q_0)^2},
\eea
with $\bm q_0=q_0\hat{\bm Q}$. Then up to the third order
\be
\alpha(\bm p+\bm q_{0})=\alpha_{0}+\alpha_{2\parallel} p_{\parallel}^2+\alpha_{2\perp} p_{\perp}^2
+\alpha_{3} p_{\parallel}^3
\ee
with following four coefficients
\bea
\alpha_0=a+\frac{q_0^2+Q^2}{2m}
-|\mathcal{J}|\sqrt{1+\lambda^4Q^2q_0^2},\\
\alpha_{2\parallel}=\frac{1}{2m}-\frac{|\mathcal{J}|\lambda^4Q^2}{2(1+\lambda^4Q^2q_0^2)^{3/2}},\\
\alpha_{2\perp}=\frac{1}{2m},\quad 
\alpha_{3}=\frac{|\mathcal{J}|\lambda^8Q^4q_0}{2(1+\lambda^4Q^2q_0^2)^{5/2}}.
\eea

\section{Appendix B: BKT diode effect and paraconductivity}
To the leading order of $\bm p$ the supercurrent is
\be
\bm J=\frac{e|\alpha_0|}{\beta}(\alpha_{2\parallel}p_{\parallel}\hat{\bm q}_0+\alpha_{2\perp}p_{\perp}\hat{\bm z}\times\hat{\bm q}_0).
\ee
To calculate the stiffness we find $\partial_{\parallel}^2\alpha=2(\alpha_{2\parallel}+3\alpha_3 p_{\parallel})$,
$\partial_{\perp}^2\alpha=2\alpha_{2\perp}$, and $\partial_{\parallel}\partial_{\perp}\alpha=0$. Thus 
\be
\det\rho\propto(\partial_{\parallel}^2\alpha)(\partial_{\perp}^2\alpha)=4\alpha_{2\perp}(\alpha_{2\parallel}+3\alpha_3 p_{\parallel}).
\ee
In other words, to the leading order of $\bm J$
\bea
\det\rho\propto \alpha_{2\parallel}\left(1+ \frac{3\alpha_3}{|\alpha_0|\alpha_{2\parallel}^2}\frac{\beta}{e}J_{\parallel}\right),\\
T_{\rm BKT}\propto\sqrt{\det\rho}\propto \sqrt{\alpha_{2\parallel}}\left(1+ \frac{3}{2}\frac{\alpha_3}{|\alpha_0|\alpha_{2\parallel}^2}\frac{\beta}{e}J_{\parallel}\right).
\eea

To calculate paraconductivity, we employ the Langevin equation to describe superconducting fluctuations
\be
\gamma\partial_{t}\Delta_{\bm q}(t)=-\alpha(\bm q-2e\bm Et)\Delta_{\bm q}(t)+f_{\bm q}(t),
\ee
where order parameter $\Delta_{\bm q}(t)$ is allowed to evolve with time $t$, $\gamma>0$ is the inverse of damping constant, and 
$f_{\bm q}(t)$ is the uncorrelated random force, which
has a Gaussian probability distribution and a temperature-dependent correlation function
\be
\langle f_{\bm q}(t)f_{\bm q'}(t')\rangle=2\gamma T\delta_{\bm q\bm q'}\delta(t-t')
\ee
due to the Einstein relation. The fluctuation will mix Cooper pairs with different momenta $\bm q$ and the current carried by damped Cooper pairs is the summation
\be\label{eq_Jt}
\bm J(t)=\int\frac{d^2\bm q}{(2\pi)^2}2e\langle |\Delta_{\bm q}(t)|^2\rangle\partial_{\bm q}\alpha,
\ee
where we obtain by solving the Langevin equation
\be
\langle |\Delta_{\bm q}(t)|^2\rangle = \frac{2T}{\gamma}\int^t_{-\infty}dt'\exp\left\{-\frac{2}{\gamma}\int_{t'}^t \alpha(\bm q-2e\bm E\tau)d\tau\right\}.
\ee

The linear paraconductivity is due to both fluctuating superconductivity $(\sigma_s=e^2\gamma T/2\pi\alpha_0)$ and normal states (denoted by $\sigma_n$).
The nonlinear paraconductivity is mainly due to fluctuating superconductivity.

\section{Appendix C: Currents}
The in-plane supercurrent density of layer $l$ is
\be
\bm J_l=\left.\frac{\partial f}{\partial\bm A_l}\right|_{\bm A_l=0}=\frac{2e}{m}{\rm Re}\left\{\psi_l^*[-i\nabla+(-1)^l\bm Q]\psi_l\right\},
\ee
and the out-of-plane supercurrent density $J_z$ reads
\be
J_z=\left.\frac{\partial f}{\partial A_z}\right|_{A_z=0}=iedJ(\psi_1^*\psi_2-\psi_2^*\psi_1),
\ee
which is nothing but the Josephson current density.

It turns out that $J_z\equiv 0$ in the orbital FF phase since $\psi_1$ and $\psi_2$ are in-phase.
In the orbital LO phase, we have
\be
J_z=2edJN_{\rm sc}\gamma\sin(2\bm q\cdot\bm r+\Theta)
\ee
where $\Theta$ is the average phase difference between two layers, and $N_{\rm sc}=\frac{1}{2}|\alpha|/{\beta}_{+}$ is the optimal Cooper pair density of the orbital LO phase. 

Then the Josephson current of the orbital LO phase is
\be
I_z=\int d^2\bm r J_z=I_0\frac{\sin(qL_{\perp})}{qL_{\perp}}\sin\Theta,
\ee
where the maximal supercurrent is $I_0=2edL_{\perp}L_{\parallel}JN_{\rm sc}$ with the sample sizes perpendicular $(L_{\perp})$ and parallel $(L_{\parallel})$ to the magnetic field direction. We have the Josephson current at average phase difference $\Theta$
\be\label{eq_JJ}
I_z=I_c\sin\Theta,\quad I_c=I_0{\rm sinc}\left(\frac{\pi\Phi}{\Phi_0}\frac{q}{Q}\right),
\ee
with sinc function ${\rm sinc}x=\sin x/x$, the magnetic flux quantum $\Phi_0=h/(2e)$ and the magnetic flux
\be\label{eq_Phi}
\Phi=BdL_{\perp}.
\ee

\section{Appendix D: Defects}
In this session, we can parameterize $\Delta_{\pm}$ as
\be
\begin{pmatrix}
    \Delta_+\\
    \Delta_-
\end{pmatrix}=
\Delta Z,\quad
Z=
e^{i\chi}
\begin{pmatrix}
    e^{-i\phi/2}\cos\frac{1}{2}\theta\\
    e^{i\phi/2}\sin\frac{1}{2}\theta
\end{pmatrix},
\ee
where $\Delta=\sqrt{|\Delta_{+}|^2+|\Delta_{-}|^2}\geq 0$, and the phases are $\chi,\phi\in[0,2\pi)$ and $\theta\in[0,\pi]$.
The two phases $\theta,\phi$ can be absorbed to a unit vector
\be
\bm n\equiv Z^{\dagger}\bm\sigma Z=(\cos\phi\sin\theta,\sin\phi\sin\theta,\cos\theta)
\ee
on the Bloch sphere. 
Two types of FF phases correspond to $\theta=0$ and $\pi$, namely two poles of the Bloch sphere; the LO phase corresponds to $\theta=\pi/2$, namely a point at the equator. Thus the symmetry group is $\mathbb{Z}_2$ for FF phase and $S^1\cong$U(1) for LO phase.
On Bloch sphere, the symmetry group $G$ is represented as
\bea
U_{\omega}&:&\chi\to\chi+\omega\\\nonumber
T_{\bm a}&:& \phi\to\phi-2\bm q\cdot\bm a\\\nonumber
I&:&\theta\to\pi-\theta,\quad\phi\to-\phi\\\nonumber
C_2\mathcal{T}&:&\theta\to\theta,\quad\phi\to-\phi
\eea
which form the subgroup $S^2\times\mathbb{Z}_2\subset G$, since the translation along $\bm q$-perpendicular direction is trivial here.

The full symmetry group of LO phase is U(1)$\times$U(1). With the superfluid stiffness $\rho_s=2\Delta^2/m$, the London free energy density can be worked out as
\bea\nonumber
f_{\rm L}&=&\frac{\rho_s}{4}\left(\nabla\chi-\frac{1}{2}\nabla\phi\right)^2+\frac{\rho_s}{4}\left(\nabla\chi+\frac{1}{2}\nabla\phi\right)^2\\\label{eq_L}
&=&\frac{\rho_s}{2}\left\{(\nabla\chi)^2+\frac{1}{4}(\nabla\phi)^2\right\}.
\eea
The topological defects in the LO phase are thus U(1) vortices with winding number $N$
\be
N=\frac{1}{2\pi}\oint d\bm r\cdot\nabla\chi\in\mathbb{Z}
\ee
and density wave dislocations with Burger's vector $\bm b$:
\be
2\bm q\cdot\bm b=\frac{1}{2\pi}\oint d\bm r\cdot\nabla\phi\chi\in\mathbb{Z}.
\ee

At tricritical point $\mathcal{T}_2$, the emergent symmetry group is SU(2) and London free energy density reads \cite{Dimi,book}
\be\label{eq_SU2}
f_{\rm L}=\frac{\rho_s}{2}\left\{\bm a^2+\frac{1}{4}(\nabla\bm n)^2\right\},
\ee
where $\nabla\chi$ becomes the generalized vector potential
\be
\bm a\equiv -iZ^{\dagger}\nabla Z=\nabla\chi -\frac{1}{2}\cos\theta\nabla\phi,
\ee
and $\nabla\phi$ evolves into the tensor $\nabla\bm n$.
In this case, the topological defects are U(1) vortices and SU(2) skyrmions, leading to the total magnetic induction
\be
B_z=\frac{\Phi_0}{2\pi}\left(\mathcal{W}+\frac{1}{2}\mathcal{Q}\right),
\ee
where flux density $\mathcal{W}$ is generated by U(1) vortices and $\mathcal{Q}$ by SU(2) skyrmions with topological invariants $W$, $Q$
\bea
\mathcal{W}=\partial_x a_y-\partial_y a_x,\quad W=\frac{1}{2\pi}\int\mathcal{W}d^2\bm r\in\mathbb{Z}\\
\mathcal{Q}=\bm n\cdot\partial_x \bm n\times\partial_y\bm n,\quad Q=\frac{1}{4\pi}\int\mathcal{Q}d^2\bm r\in\mathbb{Z}
\eea
and the total magnetic flux along out-of-plane direction is quantized according to the topological invariant $W+Q$
\be
\Phi=\int B_zd^2\bm r=(W+Q){\Phi_0}.
\ee
For a single defect, $W$ is the winding number of a U(1) vortex, and $Q$ is the Pontryagin index of a skyrmion.


\begin{thebibliography}{99}
\bibitem{FF} P. Fulde and R. A. Ferrell, Phys. Rev. \textbf{135}, A550 (1964).

\bibitem{LO} A. I. Larkin and Yu. N. Ovchinnikov, Sov. Phys. JETP \textbf{20}, 762 (1965).

\bibitem{JMLu} J. M. Lu, O. Zeliuk, I. Leermakers, Noah F. Q. Yuan, U. Zeitler, K. T. Law, and J. T. Ye, Science \textbf{350}, 1353 (2015).

\bibitem{XXi} Xiaoxiang Xi, Zefang Wang, Weiwei Zhao, Ju-Hyun Park, Kam Tuen Law, Helmuth Berger, László Forró, Jie Shan, and Kin Fai Mak, Nat. Phys. \textbf{12}, 139 (2016).

\bibitem{YuS} Yu Saito, Tsutomu Nojima, Yoshihiro Iwasa, Nature Reviews Materials \textbf{2}, 16094 (2016).

\bibitem{CXL} Chao-Xing Liu, Phys. Rev. Lett. \textbf{118}, 087001 (2017).

\bibitem{OFFLO1a} Dong Zhao, Lukas Debbeler, Matthias Kühne, Sven Fecher, Nils Gross and Jurgen Smet, Nat. Phys. \textbf{19}, 1599 (2023).

\bibitem{OFFLO1} Puhua Wan, Oleksandr Zheliuk, Noah F. Q. Yuan, Xiaoli Peng, Le Zhang, Minpeng Liang, Uli Zeitler, Steffen Wiedmann, Nigel E Hussey, Thomas T. M. Palstra, Jianting Ye, Nature \textbf{619}, 46 (2023).

\bibitem{OFFLO1b} Zongzheng Cao, Menghan Liao, Hongyi Yan, Yuying Zhu, Liguo Zhang, Kenji Watanabe, Takashi Taniguchi, Alberto F. Morpurgo, Haiwen Liu, Qi-Kun Xue, Ding Zhang, arXiv:2409.00373 (2024).

\bibitem{OFFLO1e} Xinming Zhao, Guoliang Guo, Chengyu Yan, Noah F. Q. Yuan, Chuanwen Zhao, Huai Guan, Changshuai Lan, Yihang Li, Xin Liu, Shun Wang, arXiv: 2411.08980(2024).

\bibitem{OFFLO1c} Chang-woo Cho, Kwan To Lo, Cheuk Yin Ng, Timothée T. Lortz, Abdel Rahman Allan, Mahmoud Abdel-Hafiez, Jaemun Park, Beopgil Cho, Keeseong Park, Rolf Lortz, arXiv:2312.03215 (2024).

\bibitem{OFFLO1d} F. Z. Yang, H. D. Zhang, Saswata Mandal, F. Y. Meng, G. Fabbris, A. Said, P. Mercado Lozano, A. Rajapitamahuni, E. Vescovo, C. Nelson, S. Lin, Y. Park, E. M. Clements, T. Z. Ward, H.-N. Lee, H. C. Lei, C. X. Liu, H. Miao, arXiv:2407.10352 (2024).

\bibitem{OFFLO2} Noah F. Q. Yuan, Phys. Rev. Research \textbf{5}, 043122 (2023).

\bibitem{OFFLO3} Ying-Ming Xie, K. T. Law, Phys. Rev. Lett.\textbf{131}, 016001 (2023).

\bibitem{GWQ} G.-W. Qiu and Y. Zhou, Phys. Rev. B \textbf{105}, L100506 (2022).

\bibitem{HY} Hongyi Yan, Haiwen Liu, Yi Liu, Ding Zhang, X. C. Xie, arXiv:2409.20336 (2024).

\bibitem{UN} Uddalok Nag, Jonathan Schirmer, Enrico Rossi, C.-X. Liu, J. K. Jain, arXiv:2408.00689 (2024).




\bibitem{OFFLO1q} Puhua Wan and Jianting Ye, {private conversations}.

\bibitem{Ando} F. Ando, Y. Miyasaka, T. Li, J. Ishizuka, \textit{et al.} Nature \textbf{584}, 373 (2020).

\bibitem{Chris} C. Baumgartner, L. Fuchs, A. Costa, \textit{et al.} Nature Nanotechnology \textbf{17}, 39 (2022).

\bibitem{Lorenz} L. Bauriedl, C. Bäuml, L. Fuchs, C. Baumgartner, N. Paulik, J. M. Bauer, K.-Q. Lin, J. M. Lupton, T. Taniguchi, K. Watanabe, C. Strunk and N. Paradiso, Nat. Comm. \textbf{13}, 4266 (2022).

\bibitem{Diez} J. Diez-Merida, A. Diez-Carlon, S. Y. Yang, Y.-M. Xie, X.-J. Gao, K. Watanabe, T. Taniguchi, X. Lu, K. T. Law, and Dmitri K. Efetov, Nat. Commun. \textbf{14}, 2396 (2023).

\bibitem{LinJ} Jiang-Xiazi Lin, Phum Siriviboon, Harley D. Scammell, Song Liu, Daniel Rhodes, K. Watanabe, T. Taniguchi, James Hone, Mathias S. Scheurer and J.I.A. Li, Nat. Phys. \textbf{18}, 1221 (2022). 

\bibitem{Yuan} Noah F. Q. Yuan and Liang Fu, PNAS \textbf{119} (15) e2119548119 (2022).

\bibitem{Akito} Akito Daido, Yuhei Ikeda, and Youichi Yanase, Phys. Rev. Lett. \textbf{128}, 037001 (2022).

\bibitem{James} James Jun He, Yukio Tanaka, and Naoto Nagaosa, New J. Phys. \textbf{24}, 053014 (2022).

\bibitem{Harley} Harley D. Scammell, J. I. A. Li and Mathias S. Scheurer, 2D Mater. \textbf{9}, 025027 (2022).

\bibitem{Zhai} B. Zhai, B. Li, Y. Wen, F. Wu, and J. He, Phys. Rev. B \textbf{106}, L140505 (2022).

\bibitem{Ilic} S. Ili\'c and F. S. Bergeret, Phys. Rev. Lett. \textbf{128}, 177001 (2022).

\bibitem{Marg} Margarita Davydova, Saranesh Prembabu, Liang Fu, Sci. Adv. \textbf{8}, eabo0309 (2022).

\bibitem{Banabir} Banabir Pal, Anirban Chakraborty, Pranava K. Sivakumar, \textit{et al.} Nat. Phys. \textbf{18}, 1228 (2022).

\bibitem{James1} James Jun He, Yukio Tanaka, Naoto Nagaosa, Nat. Comm. \textbf{14}, 3330 (2023).

\bibitem{JXHu} Jin-Xin Hu, Zi-Ting Sun, Ying-Ming Xie, and K. T. Law, Phys. Rev. Lett. \textbf{130}, 266003 (2023).

\bibitem{Samokhin} K. V. Samokhin, B. P. Truong, Phys. Rev. B \textbf{96}, 214501 (2017).


\bibitem{OFFLO4} Kyohei Nakamura, Akito Daido, and Youichi Yanase, Phys. Rev. B \textbf{109}, 094501 (2024).

\bibitem{Sch} A. Schmid, Phys. Rev. \textbf{180}, 527 (1969).

\bibitem{book} Boris V. Svistunov, Egor S. Babaev, Nikolay V. Prokof'ev, \textit{Superfluid States of Matter} (1st Edition), CRC press (2015).

\bibitem{Dimi} O. Dimitrova and M. V. Feigel’man, Phys. Rev. B \textbf{76}, 014522 (2007).

\bibitem{JWang} Jian Wang, {private conversations}.

\bibitem{DR} D. Rhodes, N. F. Q. Yuan, Younghun Jung, Abhinandan Antony, Hua Wang, Bumho Kim, Yu-che Chiu, Takashi Taniguchi, Kenji Watanabe, Katayun Barmak, Luis Balicas, Cory R Dean, Xiaofeng Qian, Liang Fu, Abhay N Pasupathy, James Hone, Nano letters \textbf{21} (6), 2505 (2021).

\bibitem{DR1} Apoorv Jindal, Amartyajyoti Saha, Zizhong Li, Takashi Taniguchi, Kenji Watanabe, James C. Hone, Turan Birol, Rafael M. Fernandes, Cory R. Dean, Abhay N. Pasupathy, Daniel A. Rhodes, Nature \textbf{613}, 48 (2023).

\bibitem{DR2} Zizhong Li, Apoorv Jindal, Alex Strasser, Yangchen He, Wenkai Zheng, David Graf, Takashi Taniguchi, Kenji Watanabe, Luis Balicas, Cory R. Dean, Xiaofeng Qian, Abhay N. Pasupathy, Daniel A. Rhodes, Phys. Rev. Lett. \textbf{133}, 216002 (2024).

\bibitem{KFMak} Yiyu Xia, Zhongdong Han, Kenji Watanabe, Takashi Taniguchi, Jie Shan, Kin Fai Mak, arXiv:2405.14784 (2024).

\bibitem{CoryDean} Yinjie Guo, Jordan Pack, Joshua Swann, Luke Holtzman, Matthew Cothrine, Kenji Watanabe, Takashi Taniguchi, David Mandrus, Katayun Barmak, James Hone, Andrew J. Millis, Abhay N. Pasupathy, Cory R. Dean, Nature \textbf{637}, 839 (2025).





%


\end{thebibliography}
\end{document}